\documentclass[notitlepage,a4paper,superscriptaddress,amsmath,nofootinbib,amssymb,showkeys,aps,pre]{revtex4-2}
\usepackage{graphicx}
\usepackage[usenames]{color}
\usepackage{url}
\usepackage{longtable}
\usepackage{natbib}
\usepackage{bm}
\usepackage[subnum]{cases}
\usepackage{afterpage}

\newcommand{\lambdaVent}{\lambda_{\textrm{air}}}
\newcommand{\lambdaDep}{\lambda_{\textrm{dep}}}

\newcommand{\deltat}{\delta t}
\newcommand{\texp}{\langle \tau_{\textrm{exp}} \rangle }

\newcommand{\rhoS}{\rho_{\textrm{scale}}}

\newcommand{\diPost}{d_i^{\textrm{post}}}
\newcommand{\diPre}{d_i^{\textrm{pre}}}
\newcommand{\SmallDiam}{d_1^{\textrm{post}}}
\newcommand{\LargeDiam}{d_2^{\textrm{post}}}
\newcommand{\taul}{\tau_{\textrm{lat}}}
\newcommand{\imax}{i_{\textrm{max}}}

\newcommand{\zevap}{\zeta_{\textrm{evap}}}
\newcommand{\Npath}{N_{\textrm{path}}^{(i)}}

\newcommand{\Sqss}{S_{\textrm{qss}}}
\newcommand{\Eqss}{E_{\textrm{qss}}}
\newcommand{\Iqss}{I_{\textrm{qss}}}

\newcommand{\beq}{\begin{equation}}
\newcommand{\eeq}{\end{equation}}

\begin{document}

\title{On modelling airborne infection risk}

\author{Yannis Drossinos}
\email[email address: ]{yannis.drossinos@gmail.com}
\affiliation{Thermal Hydraulics \& Multiphase Flow Laboratory,
Institute of Nuclear \& Radiological Sciences and Technology and Technology, Energy \& Safety,
National Centre for Scientific Research ``Demokritos", 15314 Agia Paraskevi, Greece}
\author{Nikolaos I. Stilianakis}
\email[email address: ]{nikolaos.stilianakis@ec.europa.eu}
\thanks{Corresponding author}
\affiliation{Joint Research Centre (JRC), European Commission,  21027 Ispra (VA), Italy}
\affiliation{Department of Biometry and Epidemiology,  University of 
Erlangen-Nuremberg,  Erlangen, Germany}

\date{\today}


\begin{abstract}

Airborne infection risk analysis is usually performed for enclosed spaces where susceptible individuals
are exposed to infectious airborne respiratory droplets by inhalation. It is usually based on
exponential, dose-response models of which a widely used variant is the Wells-Riley (WR) model.
We revisit this infection-risk estimate and extend it to the population level. We use an
epidemiological model where the mode of pathogen transmission, either airborne or contact, 
is explicitly considered. We illustrate the link between epidemiological models and the WR model.
 We argue that airborne infection quanta are, up to an
overall density, airborne infectious respiratory droplets modified by a parameter that depends on
biological properties of the pathogen, physical properties of the droplet, and behavioural parameters
of the individual. We calculate the time-dependent risk to be infected during the epidemic for two
scenarios. We show how the epidemic infection risk depends on the viral latent period and the event
time, the time infection occurs. The infection risk follows the dynamics of the infected
population. As the latency period decreases, infection risk increases. The longer a susceptible is
present in the epidemic, the higher is its risk of infection by equal exposure time to the mode of
transmission.

\end{abstract}

\maketitle

\section{Introduction}
\label{sec:Intro}

The determination of the risk of infection during an epidemic is
an important quantitative indicator that, among others,
influences decisions of public health authorities on intervention
strategies and their implementation,  including vaccine administration.
It contributes, also, to individual decisions whether to
accept recommendations of public health authorities on
social distancing, proper wearing of
face masks, and mobility restrictions. 
Estimates of the risk associated with airborne respiratory-pathogen
infection have become numerous since
the beginning of the coronavirus disease 2019 (COVID-19) pandemic.

Most airborne infection-risk analyses during the COVID-19 pandemic concentrated on
risk calculations in small, enclosed spaces within which susceptible individuals
are exposed to infectious airborne respiratory droplets by inhalation for a  brief period.
For example, the probability of infection due to the Severe Acute Respiratory
Syndrome CoronaVirus 2, (SARS-CoV-2) has been estimated in
numerous private and public
micro-environments\cite{Buonanno2020a,Buonanno2020b,Peng2022,
Poydenot2022,Jones2021,Henriques2022,Tang2022}.
The majority of these risk analyses were based on the exponential,
dose-response Wells-Riley (WR) model or its variants.

The Wells-Riley~\cite{Riley1978, RudnickMilton2003, Noakes2006} model is a deterministic 
exposure model,  based on the probabilistic airborne infection model
proposed by Wells~\cite{Wells1955}. Wells introduced
the quantum of airborne infection~\footnote{The choice of the
word quantum refers to the discrete nature of the carriers
of the airborne infection, i.e., the infectious respiratory
droplets. Wells might have been paying homage to the
great successes of quantum mechanics, universally accepted by the
time he introduced the notion of a quantum of infection.}
to be a discrete entity of the infectious dose that would, 
according to a Poisson distribution, give 
a $63.21\%$ probability of infection~\cite{Nordsiek2021} 
or, in modern
terminology,  the Infectious Dose ID$_{63.21}$. 
Riley et al.~\cite{Riley1978},  expanding on Riley (1974)~\cite{Riley1974}
and using Wells'
quantum of infection, introduced the average number of
quanta inhaled during an individual's exposure to an airborne pathogen
in an exponential dose-response model.
They, thus, proposed
a model for the risk of airborne infection in an indoor environment. 
They assumed that the micro-environment is homogeneous,
and hence infection quanta were uniformly distributed, 
and that the quantum concentration  and the
ventilation conditions were at steady state.
The resulting steady-state model 
is  commonly referred to as the Wells-Riley model.
Moreover,  they took $I$, the number of infectors, 
constant during exposure, but not so the number of susceptibles $S$,
assuming that the viral latent period,  the time between being infection 
and becoming infectious, is much longer than the exposure time,  the duration
individuals are exposed to the pathogen.
An important generalisation of the WR model was proposed by Gammaitoni and
Nucci (GN)~\cite{GN1997}. They removed the assumption of steady-state quantum
concentration to generalise it to
time-dependent quanta concentrations.

One of the characteristics of the WR model is
that it uses input from aerosol dynamics 
to estimate viral transmissibility
in, e.g., calculations of the generation rate of
the quanta of infection, and their removal rate via e.g., gravitational settling
or indoor-air ventilation.  
Human behaviour, however, is naively modeled by the lumped parameter of exposure
time. The model,  being an individual-level model and in contrast to compartmental epidemiological models,
does not consider the total population $N$. 
Instead, the enclosed-space volume $V$ determines the system
scale. 

Infection risk estimates in larger, including closed or semi-closed, populations and 
at longer, but intermediate, spatial and temporal time scales than those
investigated by micro-environmental models are equally important.  Envisioned
intermediate spatial scales are those encountered in, e.g.,
hospitals, prisons, ships, nursing homes.
Mesoscopic epidemiological models address these scales. 
The Susceptible-Infected-Recovered model with
explicit modelling of the dynamics of the pathogen carrying agent (SIR-DC model~\cite{NikosYD2010,Editorial2})
is one such model. The model considers that the pathogen-carrying agents,  the pathogen ``vector",
are either airborne infectious respiratory droplets (D) or deposited droplets (denoted by
C,  since the corresponding transmission mode is direct or indirect contact).
In modelling the dynamics of the pathogen agent the SIR-DC model 
differs from the standard SIR model where
the mode of pathogen transmission is only \emph{implicitly} 
considered.
Moreover, and contrary
to micro-environmental models, the SIR-DC
model is a population-level model.

Macroscopic models, on the other hand,  address much larger populations and much longer temporal
and spatial scales, for example country-wide 
and province scales~\cite{PGK-Greece,PGK-Mexico,NikosGreece}
or regional scales~\cite{ZoiMetaPopulation}. At such scales, 
micro-environmental dynamics are not modeled.
Instead,  the intricate dynamics of respiratory droplets and
other micro-environmental processes
are implicitly incorporated
via effective transmission rates or parameters,
via a procedure akin to coarse-grained descriptions of physical systems~\cite{Editorial2}.

Noakes et al.~\cite{Noakes2006} presented an early attempt to reconcile the WR expression
with a standard SIR compartmental epidemiological model.
We  use an extended version of the SIR-DC droplet model
to revisit the derivation
and to estimate what we shall refer to as
the epidemic infection risk, the infection
risk during an epidemic. 
We establish firmly
the connection between compartmental epidemiological models and micro-environmental risk models,
like the Wells-Riley model and its Gammaitoni-Nucci generalisation,
and the relevance of respiratory droplet dynamics.
One of the essential observations 
is that neither the GN nor the WR model considers time-dependent
changes of  the infected population.
In establishing this connection,
we elucidate the meaning
of the term \emph{quantum of airborne infection}, introduced
by William Firth Wells in his classic
1955 book~\cite{Wells1955}.  As argued in Ref.~\cite{Nardell2016},
this term has been often confusingly interpreted.

Lastly, we calculate the epidemic infection risk, the probability to
be infected, the event, at a specified time later than an
arbitrarily chosen time during the epidemic.
In essence, the question we address is: what is the probability
at time $t$ to be
infected at a later time $t+\deltat$.  The time $t$, with respect
to the beginning of the epidemic, is the time at which risk is evaluated,
and the time $t + \deltat$ is the event time.
In our numerical simulations, we calculate this probability as a function
of the pathogen latent period and
as a function of the difference between the event time and
the time the infection risk is calculated, $\deltat$, the risk time.

\section{Infection probability in compartmental epidemiological models}

The epidemic infection risk $P(t, \deltat; \texp)$, for
an average daily exposure time $\texp$, is
the probability at time $t$ from the beginning of an
epidemic to be infected at a future
time $t + \deltat$ as specified by the event, i.e., the infection.
The time difference $\deltat$  
is the time interval 
that determines how the infection risk at $t$ depends on the
evolution of the epidemic at $t+\deltat$.  We shall refer to
it as the risk time. The average daily exposure
time is an estimate of a susceptible's daily exposure to the pathogen:
it is taken to be constant during an epidemic. 
For example, in the SIR-DC model the population dynamics depend
on an average, daily exposure time which is
embedded in the transmission rate (see Supplementary Material (SM)).
In the standard SIR model the dynamics depend on
the average exposure time implicitly via, e.g., the daily number of contacts between susceptible and infected  individuals.
As the average exposure time is taken to be constant,
and the dependence of the probability on it is implicit, we shall simplify
notation and refer to the infection probability as $P(t, \deltat)$
and the number of susceptibles as $S(t)$ (instead of $S(t; \texp)$.

The epidemic infection risk expressed in terms of the number of susceptible individuals $S$
is their relative change~\cite{Beggs2003, Noakes2006}
in the period $[t, t + \deltat]$, 
\beq
P(t, \deltat)  = \frac{S(t)
- S(t + \deltat)}{S(t) } .
\label{eq:ProbInfection}
\eeq
Equation~(\ref{eq:ProbInfection}) provides the connection between epidemiological
compartmental models and infection-risk models.
Any epidemiological
model that calculates $S(t)$ can
be used to calculate the probability of infection. 
In fact, the two approximation we will consider, WR and GN, refer to
different approximate ways to calculate $S(t+ \deltat)$.

The epidemic infection risk depends explicitly on two time scale 
($t, \deltat$) and implicitly on $\texp$,
in stark contrast to WR-based risk estimates
that apparently depend on a single time scale,  the time interval a susceptible is exposed
to the pathogen.  The time $t$, the time at which risk is calculated,
may be additionally considered
in WR-models as it
determines the initial number of infectors. 
The risk time $\deltat$, i.e., difference between event and the time risk
is evaluated,
is analogous to the exposure time in WR-based model, in that
it determines the interval over which the epidemiological populations evolve, and thus the time after $t$ that
the number of susceptibles $S(t)$ is to be determined.
In the interval [$t, t+ \deltat$] the population $S(t)$ evolves according to
model dynamics, herein taken to be
the SIR-DC dynamics or 
the approximate dynamics the WR or GN models.
In WR-like models the exposure time is the time over which the
number of  susceptibles changes, and thence 
the time scale that determines infection risk.

Another important difference between
SIR-like and WR-based models is that the overall time scale of compartmental
epidemiological models is of the order of days or months, whereas the time scale
of WR-based risk analyses is of the order of hours.

\section{Susceptible-Exposed-Infected-Recovered model with transmission modes (SEIR-DC)}

\subsection {Droplet model}

As other respiratory viruses, the
SARS-CoV-2 virus exhibits a latent period. During the latent period 
$\taul = 1/\sigma$ exposed individuals are infected but not
infectious.  Accordingly, we generalize the SIR-DC  model
of infectious disease transmission via infectious respiratory droplets~\cite{NikosYD2010}
by adding an exposed population compartment $E$.
The SIR-DC model
is a population epidemiological model 
where individuals can move between the three standard epidemiological compartments of the SIR
model. In addition, it models the dynamics of the pathogen carrying agent.
Infection does not occur via direct $I \leftrightarrow S$ interaction: 
this interaction, instead, is
mediated by the infectious droplets, be they airborne or settled. 

A note is in order on the naming of the model. It was initially referred to as
SDIR~\cite{Editorial2}, a name that mixes epidemiological populations ($S,I,R$) with
the pathogen-carrying agents ($D,C$) that determine the transmission mode. 
Stimulated by Ref.~\cite{SEIR-C2023} and their SEIR-C model, we opted to separate
epidemiological populations from transmission modes. Infectious respiratory droplet that
are responsible for non-contact airborne transmission~\cite{Dichotomy} are denoted by $D$ (for droplets),
whereas settled droplets that are responsible for contact transmission, direct or indirect,
are denoted by $C$ (for contact). 

\begin{figure}[htbp]
\begin{center}
\includegraphics[scale=0.30]{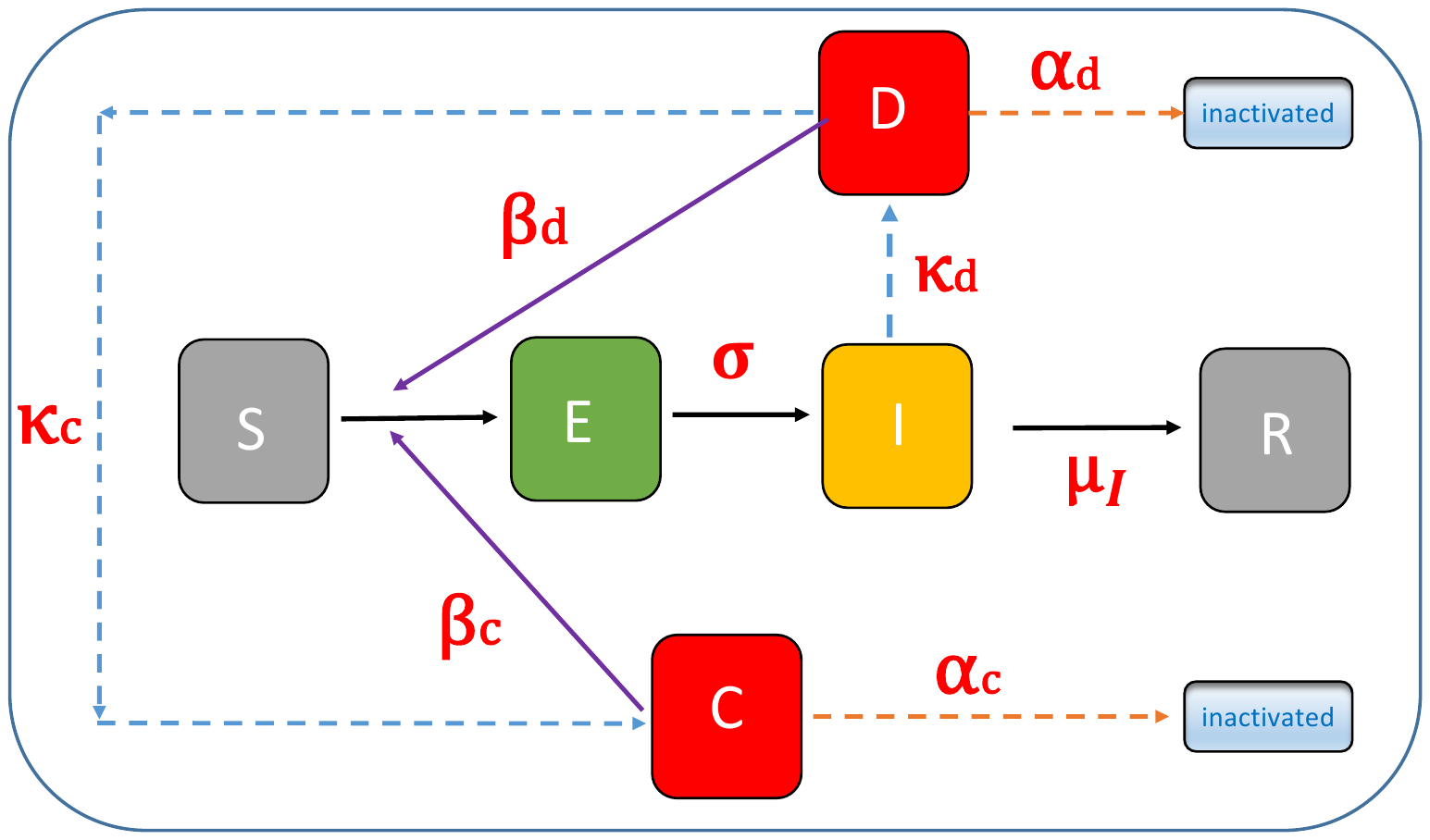}
\caption{Schematic diagram of the Susceptible-Exposed-Infected-Recovered 
model with transmission modes, SEIR-DC (based
on a  figure of Ref.~\cite{Editorial2}).  Infectious droplet compartments are denoted by $D_i$, airborne droplets, and $C_i$,
settled droplets. Superscripts ($d,c$) denote (airborne, settled) droplets, the subscript $i$ refers
to droplets with post evaporation diameter $\diPost$. Infection transmission rates are denoted by $\beta_i^{d,c}$,
droplet generation rates by $\kappa_i^{d,c}$, and removal rates by $\alpha_i^{d,c}$. The latent period
is $\taul = 1/\sigma$ and the infection recovery rate $\mu_I$.}
\label{fig:SIR-DS}
\end{center}
\end{figure}

Bazant and Bush~\cite{Bazant2021} also included the exposed population
to connect the SEIR to the WR model. They, as we do herein, also considered
cases of short and long latent periods of the pathogen. 
Our
approaches differ, however, in how the
epidemiological population compartments are coupled to the infectious
respiratory droplets.
The SEIR-DC model is defined by the following 
set of coupled ordinary differential equations (ODEs)
\begin{subequations}
\begin{eqnarray}
\frac{dS}{dt} & =  & -  \sum_{i=1}^{i=\imax} \Big ( \frac{\beta_i^d}{N} D_i S  + \frac{\beta_i^c}{N} C_i S \Big ) ,    \label{eq:Sequation} \\
\frac{dE}{dt} & =  & - \frac{dS}{dt} -\sigma E , \label{eq:Eequation} \\
\frac{dI}{dt} & = & \sigma E - \mu_I I ,  \label{eq:Iequation} \\
\frac{dD_i}{dt} & = &  \kappa_i^d I - \alpha_i^d D_i  , \quad \mbox{for} \quad i =1,2 \ldots \imax  \label{eq:Dequation} ,\\
\frac{dC_i}{dt} & = &  \kappa_i^c D - \alpha_i^c C_i , \quad \mbox{for} \quad i =1,2 \ldots \imax  \label{eq:Cequation} .
\end{eqnarray}
\label{eq:SEDIR}
\end{subequations}
We do not show the equation
for the recovered compartment $R$ since the total
population $S+I+R=N$ is constant representing a closed population.
A schematic diagram  of the model is shown in Fig.~\ref{fig:SIR-DS}.

The number of infectious airborne droplets of 
post-evaporation diameter $\diPost$ is denoted
by $D_i$ (number), and that of settled droplets by $C_i$ (number),
cf.  SM for a discussion of droplet evaporation
and associated droplet diameters.
The number of droplet classes is $\imax$. The rate of transition from the
exposed compartment $E$ to the infected compartment $I$ is denoted
by $\sigma$. 
The infection recovery rate $I \rightarrow R$ is $\mu_I$.
Superscripts denote airborne droplet ($d$) and settled ($c$),
and the subscript $i$ denotes the droplet class specified by the
post-evaporation diameter $d_i^{\textrm{post}}$.
The transmission rate per infectious,  airborne respiratory droplet that has been inhaled and
deposited in the respiratory tract of a susceptible is denoted by $\beta_i^d$ (inverse time), 
whereas that of an infectious settled droplet
transferred to facial membranes is denoted by $\beta_i^c$ (inverse time).

The airborne droplet generation rate per infected individual (by 
normal oro-nasal activities-e.g., speaking, laughing, breathing- or by
violent expiratory events - sneezing, coughing-) is
$\kappa_i^d$ (number/time) and the corresponding 
airborne droplet removal rate is $\alpha_i^d$ (number/time),
the later including  droplet removal by ventilation (if present).
Settled droplets may be generated either via direct generation by an infected individual and deposition on facial mucous tissues or
via deposition of airborne droplets. Direct deposition would introduce an
additional generation term in Eq.~(\ref{eq:Cequation}) proportional to the number
of infected individuals, similar to the generation term in the airborne-droplet
equation, Eq.~(\ref{eq:Dequation}).  In this version of the model we neglect
this mechanism. Instead, settled droplets are generated
via deposition of airborne droplets, and specifically solely by
gravitational settling. Hence the generation rate $\kappa_i^c = \theta_i(\diPost)$ (number/time) 
with $\theta$ the gravitational settling rate in still air.
The corresponding 
settled droplet removal rate is $\alpha_i^c$ (number/time).

We present expressions for the transmission $\beta_i^{c,d}$ and removal
$\alpha_i^{c,d}$ rates, along with justifications for our choices, in SM.
We remark that the transmission and removal rates are \emph{derived} quantities.
In addition, both transmission rates depend (linearly as we argue in
SM) on the average exposure time $\texp$. 
The SIR-DC basic reproduction number is~\cite{NikosYD2010,Editorial2}
\beq
R_0^{\textrm{SIR-DC}} = \sum_{i=1}^{i = \imax} \Big ( \frac{\beta_i^d \kappa_i^d}{\alpha_i^d \mu_I}
+ \frac{\beta_i^c \kappa_i^c}{\alpha_i^c \mu_I} \Big ) .
\label{eq:R0SDIR}
\eeq
Equation~(\ref{eq:R0SDIR}) also gives the SEIR-DC basic reproduction number, see, for example, Ref.~\cite{VandenDriessche}.

\subsection{Infectious quanta: Gammaitoni-Nucci approximation}
\label{section:GN}

We limit the droplet classes 
to a single airborne droplet class $D_1$ (SEIR-D) as
the original GN model considered only one droplet diameter.
It can be shown, for example by integrating the infected
population Eq.~(\ref{eq:Iequation}), that 
if $\sigma \deltat \ll 1$,  latent period much greater than the
risk time (the time the epidemiological populations evolve to calculate
the infection probability),
and $\mu_I \deltat  \ll 1$, infectiousness period much greater than
the risk time $\deltat$, then $dI/dt|_{t=t_0} =0$.
Hence,  under these limits,
the number of infected at  time $t$, here taken to be the initial time $t_0$,  
may be considered to be constant and denoted as $I_0$.  If, in addition, we
disregard the equation for the
exposed population Eq.~(\ref{eq:Eequation}), which is irrelevant over the
risk time $\deltat$ for the time-development
of the infection, $\sigma \deltat << 1$ (the number of $E$ increases, but not that of $I$),
the SEIR-D model reduces to
\begin{subequations}
\begin{eqnarray}
\frac{dS}{dt}  & = & - \frac{\beta_1^d}{N} D_1 S , \label{eq:limit1} \\
\frac{dD_1}{dt}  & = &  \kappa_1^d I_0- \alpha_1^d D_1 . \label{eq:limit2}
\end{eqnarray}
\label{eq:GNlimit}
\end{subequations}
The system of Eqs.~(\ref{eq:limit1},\ref{eq:limit2}) can be compared to
the GN equations~(\ref{eq:GN})
for the rate of change of the number of susceptibles
and total number of quanta of infection $Q$ in the enclosed space.
The GN equations
expressed in our notation read
\begin{subequations}
\begin{eqnarray}
\frac{dS}{dt} & = & - \frac{B}{V} Q S , \label{eq:GN1} \\
\frac{dQ}{dt} & = & q I_0 - \lambdaVent Q , \label{eq:GN2}
\end{eqnarray}
\label{eq:GN}
\end{subequations}
where $q$ is the quantum generation rate per infectious individual (quanta/sec), see also~\cite{Noakes2006},
 $B$ is the breathing rate (m$^3$/sec), and $V$ is the
space volume (m$^3$). 
The parameter $\lambdaVent$ is the
quantum removal rate 
which in the initial formulation of the model was taken to be
the ventilation rate in air exchanges per hour~\cite{GN1997}.
Since then,  it has been expanded to include 
the rate of pathogen inactivation, droplet surface deposition, inactivation due to UV irradiation, 
filter penetration, mask efficiency, etc. (see also the
droplet removal rates $\alpha_1^d$ used in this work and summarized in SM).

The analytical solution of Eq.~(\ref{eq:GN2}) is
\beq
Q(t) =   \frac{qI_0}{\lambdaVent}  + \left ( Q_0 - \frac{qI_0}{\lambdaVent} \right ) \exp ( - \lambdaVent \deltat ) ,
\label{eq:QuantaTime-GN}
\eeq
where $Q_0$ is the initial (at time $t=t_0$) total concentration of the infection quanta in the enclosed space.

Even though the two sets of equations Eqs.~(\ref{eq:GNlimit}) and (\ref{eq:GN})
are formally equivalent, their interpretation and the time scales chosen to determine
infection risk differ.  In our numerical simulations we use 
typical time scales associated with compartmental epidemiological models, of the order
of months.  WR-based models in micro-environments,
instead, use considerably shorten time scales, of the order of hours.

In Section~\ref{sec:Numerics},  we use the GN equations to
approximate the dynamics of the number of susceptibles $S(t)$,
and subsequently the infection risk
according to Eq.~(\ref{eq:ProbInfection}). 
As the droplet equation  Eq.~(\ref{eq:limit2}),
and subsequently the susceptible equation Eq.~(\ref{eq:limit1}), may be solved analytically,
in our numerical simulations we used the analytical solutions.

\subsubsection{What are infection quanta?}

The comparison of Eqs.~(\ref{eq:GNlimit}) and (\ref{eq:GN}) provides insights
on the differences and formal similarities of the SEIR-D and GN models. 
Let the number of quanta of infection  be proportional to the number of infectious
respiratory droplets
\beq
Q = \xi D_1 ,
\eeq
and the transmission
rate proportional to the breathing rate,  $\beta_1^d = B \tilde{\beta}_1^d$,
as argued in SM. Moreover,  for the purposes of this comparison,
consider indoor-air ventilation
the only droplet or quantum removal process, $a_1^d = \lambdaVent$.
Their substitution into Eqs.~(\ref{eq:GNlimit}), and a mapping of the resulting equations to Eqs.~(\ref{eq:GN}) determines
the conversion factor $\xi$ to be
\beq
\xi = \frac{\beta_1^d}{B} \,  \frac{V}{N} \equiv \tilde{\beta}_1^d \rhoS ,
\label{eq:xi}
\eeq
where the last equation defines the scaling density $\rhoS = V/N$.  Hence, in this model
infection quanta, up to an overall scaling density, are infectious respiratory droplets modified by $\tilde{\beta}_1^d$, a parameter
that includes the probability of infection of a lung-deposited pathogen,
number of pathogens in a droplet, lung-deposition probability,  and average exposure time, cf.  SM.
The combination of these factors converts infectious airborne droplets to infection quanta.
Their generation $q$ is similarly related to the respiratory droplet generation
rate via $q = \kappa_1^d \xi$.

The mapping of the two models also manifests the different inherent
system scales:
the extensive variable,
namely the variable that scales linearly with the size of the system, 
is the volume of the enclosed space in the GN model, whereas it becomes
the total population $N$ in the SEIR-D model.  The scaling density $\rhoS$
implements the transition from a microscopic models, which depends on the enclosed-space volume $V$, to
a mesoscopic epidemiological model,  which depends on the total population $N$.
This scaling is reminiscent of the scaling proposed in Ref.~\cite{PGK-Greece} to transition
from an ODE to a PDE epidemiological model.

Care should be exercised in interpreting $\rhoS$: if $V$ is taken to refer to a mesoscopic volume, then
the GN model is essentially extended to much greater scales.  If $N$ is taken to be
the number of individuals in an enclosed, micro-environment
the SIR-D model is restricted to smaller scales; however, in that case it may not
be considered a proper compartmental epidemiological model. These consideration have important
repercussion on the choice of model parameters and risk times in micro or mesoscale models.

\subsection{Wells-Riley approximation}

Reference~\cite{MargueriteEpidemics}
considered  analytically the very common limit
where the duration of infectiousness of
an infected individual $T_I = 1/\mu_I$ is significantly
longer than the lifespan of the airborne pathogen $T_p = 1/\alpha_1^d$,
i.e., when  $\rho_1 \equiv \mu_I/\alpha_1^d = T_p /T_I \ll 1$.  
For appropriately chosen non-dimensional variables~\cite{MargueriteEpidemics} the quasi steady-state limit
is defined as  $\rho_1 d\tilde{D}_1/d \tilde{t} =0 $ which implies $\tilde{D}_{1, \textrm{qss}} = \tilde{I}$, or
in terms of the original variables
$D_{1,\textrm{qss}} = (\kappa_1^d/\alpha_1^d) I$.
Note that the quasi steady-state condition does \emph{not} imply that the number of
infected individuals is constant,  $dI/dt|_{\textrm{qss}} \neq 0$, that is
$\Iqss$ is \emph{time dependent}.

The substitution of the steady-state ($I, D_1$) relationship in the original equations Eqs.~(\ref{eq:SEDIR}) gives the
quasi steady-state limit of SEIR-D,
\begin{subequations}
\begin{eqnarray}
\frac{d\Sqss}{dt}  & = & -  \frac{\beta_1^d \kappa_1^d}{\alpha_1^d N} \Iqss \Sqss ,  \label{eq:Sqss}\\
\frac{d\Eqss}{dt}  & = &  \frac{\beta_1^d \kappa_1^d}{\alpha_1^d N} \Iqss \Sqss- \sigma \Eqss ,  \label{eq:Eqss} \\
\frac{d\Iqss}{dt}  & = & \sigma \Eqss - \mu_I \Iqss  .  \label{eq:Iqss} 
\label{eq:SEDIR-qss}
\end{eqnarray}
\end{subequations}
In the quasi steady-state limit the dependence on
the number of infectious droplets $D_1(t)$ disappears.

As before,  in the previously considered double limit,
$\sigma \deltat, \mu_I \deltat \ll 1$,
we can neglect the equation for the exposed population, Eq.~(\ref{eq:Eqss}),
and  take the number of infected
individuals constant, $\Iqss = I_0$.  
The analytical solution of the resulting model 
leads directly to
the WR approximation of the SEIR-D model
\beq
P_{\textrm{WR}}^{\textrm{SEIR-D}}  (t, \deltat) = 1 - 
\exp \Big ( - \frac{\beta_1^d \kappa_1^d}{\alpha_1^d N} I_0 \deltat \Big ) .
\label{eq:SEDIR-WR}
\eeq
Hence, the WR equation is obtained from
the quasi steady-state SEIR-D equations in the triple limit of latent period
and infectiousness period  longer than the time scale of observation
and $\rho_1 \ll 1$.  Substitution of 
$k_1^d = q/\xi$, along with Eq.~(\ref{eq:xi}), and $\alpha_1^d = \lambdaVent$ in Eq.~(\ref{eq:SEDIR-WR})
leads to the WR infection probability as usually written.
%

Of course, the WR approximation to the GN model may be easily obtained by
setting $dQ/dt =0$ in Eq.~(\ref{eq:GN2}). 
The steady-state quantum concentration, then, becomes
$Q_{ss} = qI_0/(\lambdaVent)$, leading via Eq.~(\ref{eq:GN1})
to the number of susceptibles
and, thus, to the WR infection risk. 
However, the alternative derivation for the WR approximation
we presented in terms of the
quasi steady-state solution of the SEIR-D model specifies
the region of validity of the approximation,
instead of arbitrarily setting $dQ/dt =0$.

\section{Numerical results}
\label{sec:Numerics}

We performed numerical simulations of the SEIR-D model, Eqs.~(\ref{eq:SEDIR}), to investigate the effect of
the event time $t + \deltat$, via $\deltat$, and of the latent period $\taul$ on
epidemic infection risk.  As mentioned, we address the question of what is
the probability at time $t$,
any arbitrarily chosen time during the epidemic, to be infected at a
future event time $t + \deltat$,
given the evolution of the pandemic till $t+\deltat$.
We also investigate numerically and analytically the 
validity of the GN, Eqs.~(\ref{eq:GNlimit}),  and WR, Eq.~(\ref{eq:SEDIR-WR}),
approximations to the SIR-D population dynamics.  
\begin{figure}[htbp]
\begin{center}
\includegraphics[angle=-90,scale=0.30]{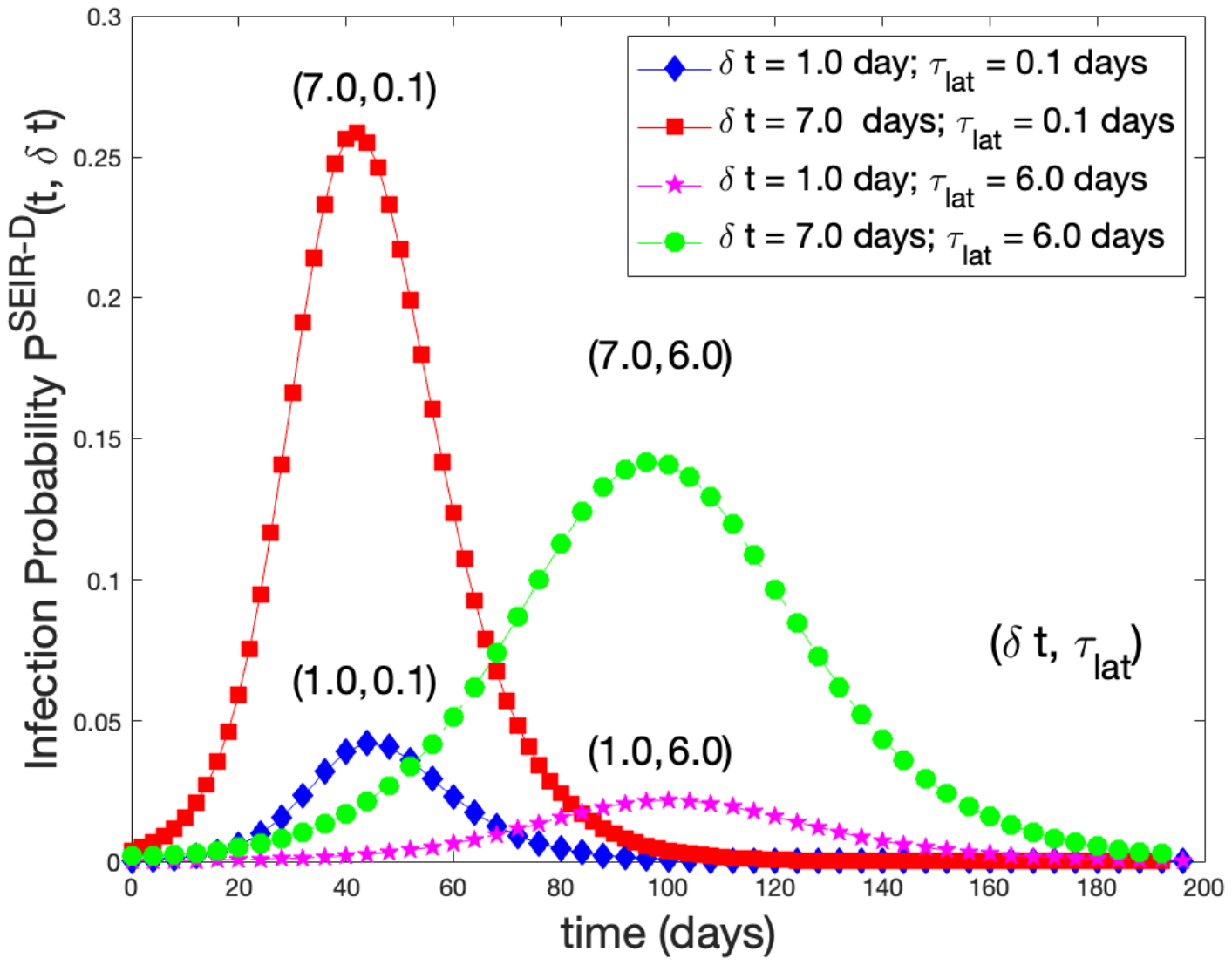}
\includegraphics[angle=-90,scale=0.30]{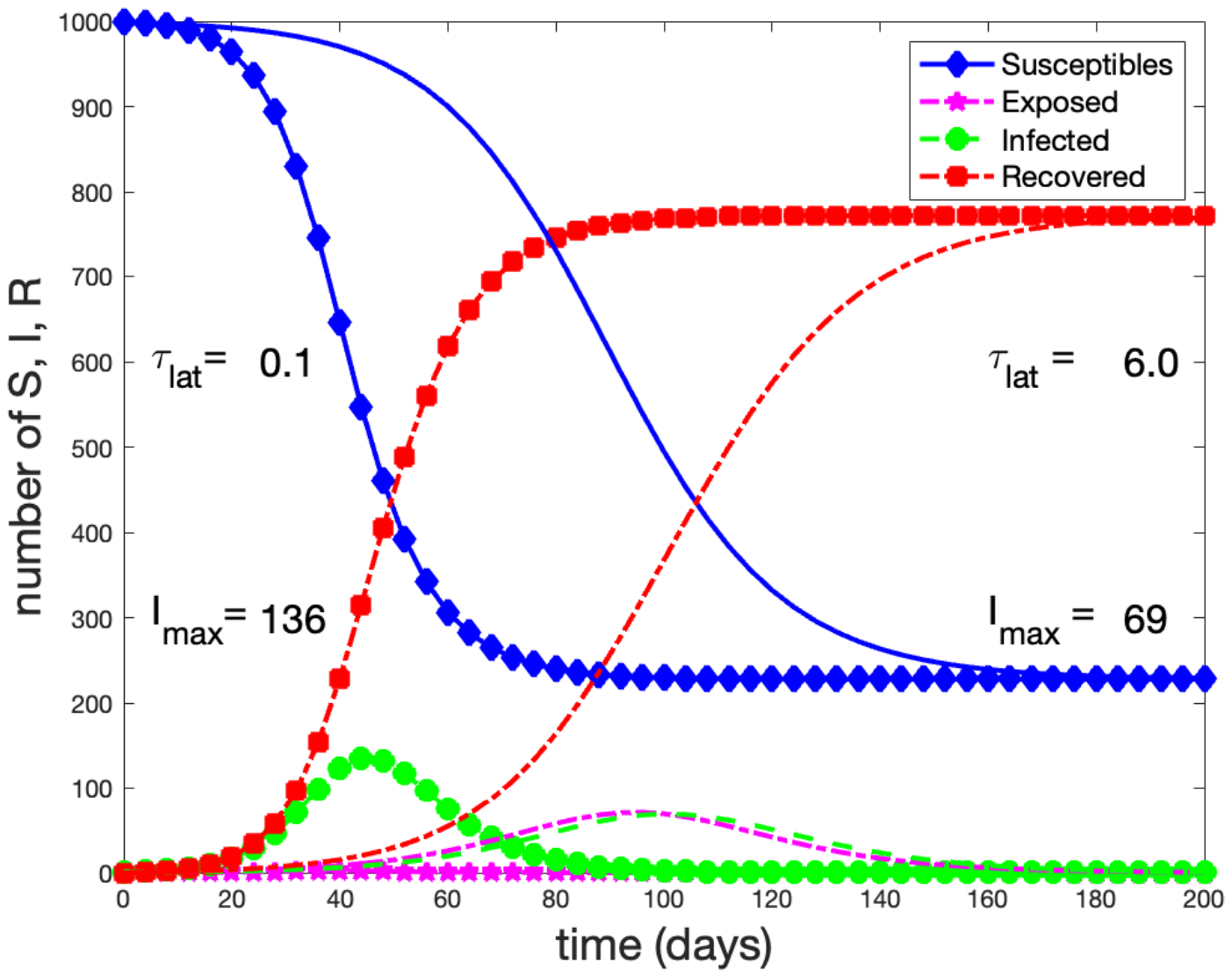}
\caption{Left: Epidemic infection probability, i.e. , the
probability at $t$ to be infected at a future time $t+\deltat$, according to the SEIR-D model. 
Curves were calculated for two risk times ($\deltat = 1, 7$ days) and two
latent periods ($\taul = 0.1, 6$ days).  Two airborne-droplet classes were considered,
($\diPost = 2.05, 82.13$ $\mu$m ($i=1,2$)), susceptible-infectious droplet encounters
per day were taken to be $c=18$,  exposure time for each $S \leftrightarrow D_i$ ($i=1,2$),
encounter $\tau_{d_1} = 25$ min and
$\tau_{d_2} = 1$ min,  leading to a total daily susceptible-infectious droplet
average exposure time of $\texp = 7.8$ hours (per day).  The ventilation rate
was taken to be $\lambdaVent = 0.2$ air exchanges per hour, a typical
value for an Italian building~\cite{Buonanno2020a}. Total
population $N = 1000$. Right: Corresponding
dynamics of the two epidemics. 
The left curves (filled symbols) correspond to the short
latent period, $\taul = 0.10$ days,  with $I$ peaking
at $t \approx 43$ days,  and no discernible exposed population. The right
curves (lines, no symbols) show the epidemic for the long latent
period $\taul = 6$ days, with $I$ peaking at $t \approx 96$ days, and an appreciable exposed population.}
\label{fig:ProbInfect-Epidemic}
\end{center}
\end{figure}

For the simulations we used parameters related to the COVID-19 pandemic,
e.g., individual behaviour characteristics in addition to physico-chemical
and biological properties of the SARS-CoV-2 virus. 
We note, though,  that we do not attempt to reproduce a COVID-19 scenario
as in our attempt to present the minimal model that reduces to the GN or WR models
we do not consider the asymptomatic stage of the disease.  

We used two airborne droplet classes of post-evaporation
diameter $\diPost = 2.05, 82.13$  $\mu$m ($i=1,2$).  As generally accepted~\cite{Santarpia2022},
the pathogen concentration was taken to be droplet-size dependent.
We opted to limit the airborne
droplet classes to two and not to simulate settled
droplets to render easier the interpretation of our results: 
either condition can be easily relaxed.
The evaporation factor~\cite{Editorial2},
$\diPost = z_{\textrm{evap}} d_i^{\textrm{pre}}$,
was set to $z_{\textrm{evap}} = 0.40$.  Airborne droplet generation rates
were taken to correspond to speaking.  A complete list of
model parameters is presented in SM.

Individual behaviour determines a number of model parameters.
We considered the contact rate,
the number of susceptible-infected individual encounters, to
be $c=18$ per day~\cite{Sypsa2021}.  The exposure time of a susceptible
with an infectious droplet,i.e.,  the breathing time during a
$S \leftrightarrow I$ encounter,
was taken to depend on the droplet size:
$\tau_{d_1} = 25$ min and $\tau_{d_2} = 1$ min.  Thus,  the average exposure
time per day of a single susceptible is $c \times (\tau_{d_1}+\tau_{d_2}) = 7.8$ hours per day.

Figure~\ref{fig:ProbInfect-Epidemic} summarizes the main results of four
simulations to determine the probability at
time $t$ that infection occurs at the event time
$t + \deltat$.
We used two latent periods $\taul = 0.1$ days (short) and $\taul = 6.0$ days
(long), along with a short $\delta t= 1.0$ day and a long relative
time interval $\deltat = 7$ days.
The left panel shows the calculated infection probabilities for each scenario. Two groups of curves may be identified:
for the short latent period the infection probability
peaks at about $t_{\textrm{peak}} \approx 43$ days,  whereas for the long latent
period the peak occurs at $t_{\textrm{peak}} \approx 96$. Within each group of curves,
infection risk increases with increasing risk time.

The qualitative behaviour of infection risk may be understood by considering the dynamics
of the epidemic,  described by the time-dependent number of $S,E,I$ and $R$ shown
in the right panel of Fig.~\ref{fig:ProbInfect-Epidemic}.
The four curves on the left (filled symbols) correspond to the short latent period,
whereas those on the right (no symbols) to the long latent period. 
We also present the maximum number of infected individuals $I_{\textrm{max}}$
for each epidemic. For the short-latent period epidemic, the number of exposed individuals is
very small,  not discernible on the figure, whereas for the long-latent period epidemic the
number of exposed individuals is comparable to the number of infected. In fact,
before the $I$ maximum, $E >I$, whereas afterwards $I>E$. Even though not discernible,
the number of exposed individuals $E$ peaks earlier than the number of infected $I$.

We find that 
infection risk follows the time-dependent behaviour, the dynamics,
of the infected individuals $I$.
As the number of infected increases, infection risk increases, and vice versa.
As the latency period decreases infection risk increases,
since the number of infected increases.
In addition, for the same pathogen (latent period constant)
infection risk increases with
increasing risk time $\deltat$, i.e., with
increasing difference between event time and time $t$, the time
risk is evaluated.
 In fact, this is a general result: the longer 
 the epidemic evolves and the longer a susceptible
 is present in it, the more likely the individual is to be infected.

\begin{figure}[htbp]
\begin{center}
\includegraphics[angle=-90,scale=0.35]{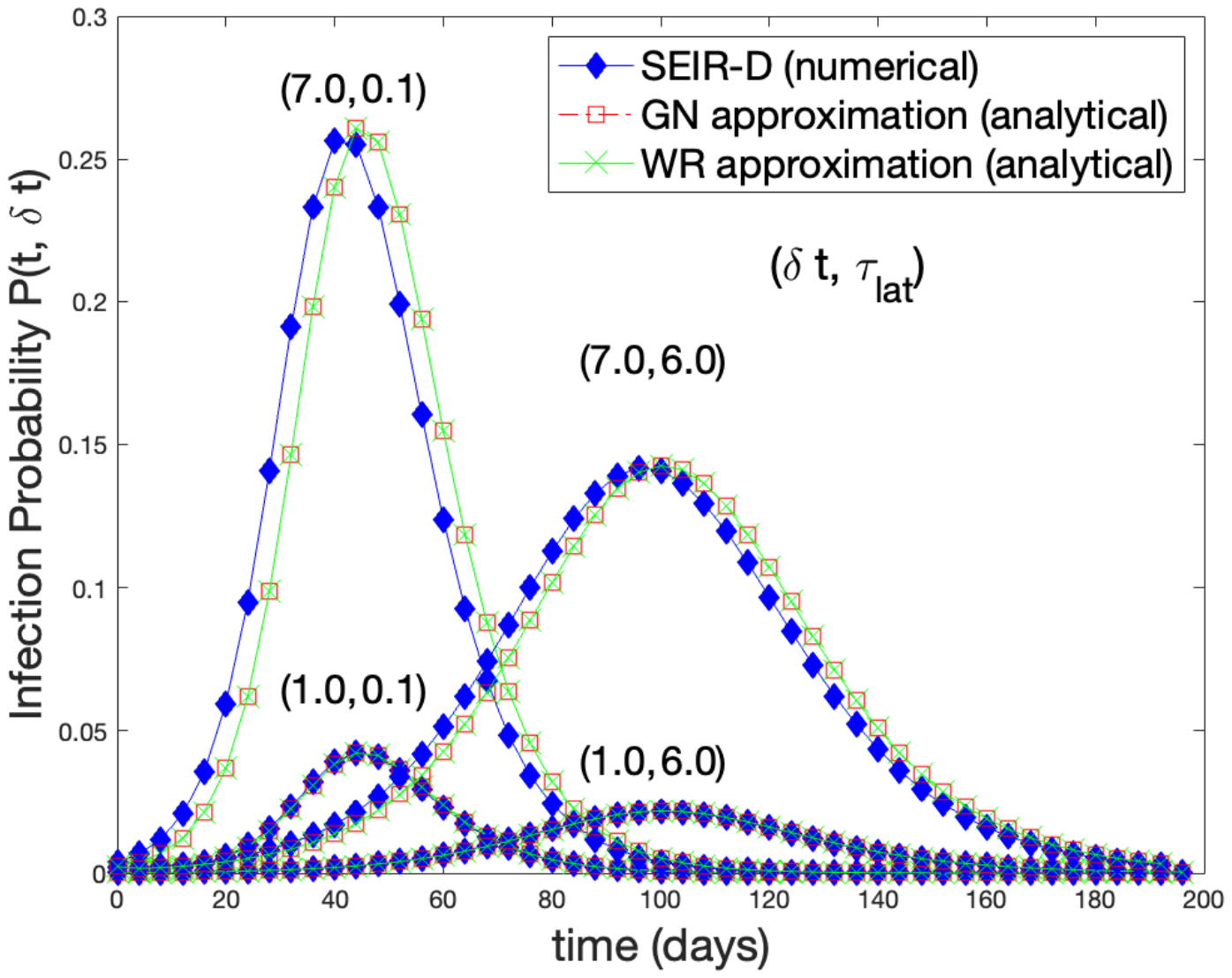}
\caption{Epidemic infection probability calculated via the SEIR-D
model and its GN, 
and WR approximations,
Eq.~(\ref{eq:SEDIR-WR}). The four ordered pairs associated
with each graph triplet are specified by ($\deltat, \taul)$. For
all the simulations the GN and WR limits were identical: they differed 
from the SEIR-D model predictions only for the long  risk time ($\deltat = 7$ days). See the main text for the explanation.}
\label{fig:SEDIR-Limits}
\end{center}
\end{figure}

The validity of the GN and WR approximations 
to the SEIR-D dynamics 
is investigated
numerically in Fig.~\ref{fig:SEDIR-Limits}.  Four groups of curves are shown, each
corresponding to the ordered pair ($\deltat, \taul)$.  For each pair choice,
we plot the SEIR-D infection risk 
calculated via the numerical solutions of Eqs. ~(\ref{eq:SEDIR}) (filled blue diamonds),
infection risk calculated via the GN approximation and described in Section~\ref{section:GN} 
(square, unfilled symbols), and via the WR approximation $P_{\textrm{WR}}^{\textrm{SEIR-D}}$ as calculated
via Eq.~(\ref{eq:SEDIR-WR}) (cross, continuous line).

Two observations are in order. For the epidemics considered, the GN and WR limits
are identical.  Whether the two limits would differ depends on the airborne droplet
removal rate $\alpha_1^d$ (and hence on the
dimensionless parameter $\rho_1 = \mu_I/\alpha_1^d$). ~\footnote{We restrict  the analysis to a single airborne droplet class, for simplicity: the
arguments are easily generalized.}
The importance of the removal rate
is apparent from the analytical solution of the droplet equation
Eq.~(\ref{eq:limit2}).
Its time-dependent part, which is formally identical to the second
term of Eq.~(\ref{eq:QuantaTime-GN}),
determines the difference between the steady-state
and non steady-state model.  It vanishes as $\alpha_1^d \deltat \gg 1$,
a condition satisfied for all cases considered.
The same observation holds for the GN-WR 
comparison. 
If ventilation is the dominant aerosol removal process,
for  $\lambdaVent \delta t \gg 1 $ the two models become identical.  Hence, for high ventilation rates
the difference between the steady and non steady-state quantum concentration models
decreases or even vanishes. 

The other observation is that for the short risk time $\deltat = 1$ day all three
calculations predict the same infection risk, irrespective of the viral latent period. 
The calculated risks differ for the long risk time, the difference increasing 
with the latent period decreasing, 
i.e., as $\taul \rightarrow 0$. 
For short latent periods the time dependence of the infected $I(t)$
can not be neglected.
As the number of $I$ increases, i.e., before the maximum of
the number of infected,  the GN and WR approximate dynamics underestimate
infection risk,
whereas the opposite holds when $dI/dt < 0$.  
%
This arises because when the number of $I$ increases more infectious droplets
are generated than predicted for a constant $I_0$ leading to a larger infection
probability, and vice versa.

In an attempt to investigate the GN-WR difference we considered an extreme
case of the SEIR-D model by shortening the model time scales from days to hours.
The calculated infection risk,  not shown, behaved as described in the previous paragraphs, 
confirming the initial observation that even for short risk times, e.g., $\deltat = 12$ hours, the
condition $\alpha_1^d \deltat \gg 1$ remained valid.  We note that the GN and WR models
may, however, differ in micro-environmental simulations if the necessary condition,
e.g., $\alpha_1^d \deltat \ll 1$ are satisfied.

\section{Discussion}

We presented a model to calculate epidemic infection risk,
namely the probability that at an arbitrarily chosen time $t$
during the epidemic infection occurs at a future event time $t + \deltat$.
The SEIR-DC model consists of the standard epidemiological populations
of the Susceptible-Exposed-Infected-Recovered (SIR) populations 
coupled to the dynamics of the pathogen-carrying agent,
and hence to the pathogen transmission mode.
The pathogen-carrying agent is taken to be either airborne infectious respiratory droplets (D),
as in the case of airborne infections such as COVID-19 or influenza,
or settled droplets responsible for (direct or indirect) contact transmission (C).
The model 
provides a connection between SEIR-like 
epidemiological models and infection-risk models 
based on Wells-Riley~\cite{Riley1978} models,
including the non-steady state generalisation proposed by Gammaitoni
and Nucci (GN)~\cite{GN1997}.
In fact, the SEIR-DC model may be viewed as a generalisation of
the GN model to the population level for arbitrary viral latent periods.
We emphasized the importance of system scales, since both the GN and WR
models are individual-level models that describe infection risk
in enclosed micro-environments.  SEIR-like models instead are population-level models.

We argued that for long latent periods of the pathogen the SEIR-DC model reduces to a set of
equations that are reminiscent of the GN equations. Their mapping 
identified infection quanta as infectious respiratory particles modified
by a scaling density and, more importantly, by a combination of parameters that include biological properties
of the pathogen (size-dependent pathogen droplet concentration, probability
of infection due to a deposited infectious droplet), physical properties (lung-deposition
probability), and behavioural properties (exposure time).
We noted that the SEIR-DC epidemiological model
depends on the total population $N$, whereas both
the WR and GN models consider much smaller scales in terms of the enclosed volume $V$.
We identified the scaling density as the factor to transition from one class of models
to the other, and we discussed how this density allows a
generalisation of micro-environmental models.

We performed numerical simulations of two scenarios
for an epidemic specified by a short and a long latent period
and driven by two classes of airborne infectious droplets.
Model parameter were based on properties of the SARS-CoV-2 virus, even
though we do not claim to model specifically the SARS-CoV-2 transmission dynamics
with all of their characteristics. However, the
SARS-CoV-2 transmission dynamics reflect those of a range of airborne infections such as influenza.

Our numerical simulations determine the risk at any
time $t$ to be infected at
a future event time $t+\deltat$,
as determined from the evolution of
the epidemic in the time interval [$t,  t + \deltat$]. 
We were particularly interested
in the dependence of the calculated infection probability on the event time
$t+\deltat$ via $\deltat$.
We found that infection risk follows the dynamics
of the infected population: as the number of infected increases
so does the risk, and vice versa.
As the latency period decreases, infection risk increase.
Moreover, as the difference between the event and the
time at which risk is evaluated increases, i.e., as the risk time increases,
infection risk increases.
Equivalently, the longer the epidemic evolves, 
the longer an individual is present in an epidemic,
the more likely that the individual gets infected, and hence infection risk
increases.

The WR and GN approximations of the SEIR-D dynamics reproduced
accurately calculated infection risk. 
Differences arose for large time intervals $\deltat$ ($\deltat = 7$ days),
increasing with decreasing viral latent period.
We remarked that the WR and GN approximate forms of the epidemic infection risk
were almost identical for all our simulations. This arises when the droplet removal
rate $\alpha_1^d$ is much greater than the inverse risk time,
i.e., $\alpha_1^d \deltat \gg 1$. In fact, this is a general result
suggesting that with increasing droplet removal rates,
for example via increased ventilation rate, 
the WR-calculated airborne infection risk with a steady-state quantum
concentration provides an excellent approximation to the GN-calculated
infection risk with non-steady-state quantum concentrations.

The comparative analysis presented here bridges the gap and provides the missing links in the
mathematical relationship between individual infection risk models and associated population based models. The
corresponding insights allow for a more nuanced epidemiological interpretation of infectious disease outbreaks. 

\begin{acknowledgments}
YD would like to thank the PEACoG (Physical Epidemiology Amherst Covid Group)
members for their many insightful discussions and helpful comments over the last years.
We thank Marguerite Robinson for discussions during  the initial stages of our work,
and Vladimir M.  Veliov for comments on the connection between the SIR-D and WR models.
The views expressed are purely those of the authors and may not
in any circumstances be regarded as stating an official position of the European Commission.
\end{acknowledgments}

\section*{Data availability statement}

The study has no original data. The numerical results of this study are available within this paper. 
The parameter values and their justification are in the Electronic Supplementary Material of this paper.

\section*{Author contributions}
Y.D. designed research, performed research, analyzed the results and wrote the article; N.I.S. designed research, analyzed the results, and wrote the article.

\section*{Competing interest}
The authors declare no competing interest

\section*{Funding}
The research was performed with institutional support only. 

\clearpage

\appendix

\section{Supplementary Material}

\subsection{Susceptible-Exposed-Infected-Recovered with transmission
modes (SEIR-DC): \\
Model parameters}

The droplet population compartments $D_i$, number of airborne droplets, and $C_i$,
number of settled droplets,
are identified by the droplet diameter. Respiratory
droplets are generated in the respiratory tract under conditions of 100\% relative
humidity and approximately 37$^o$ C degrees.  Upon expulsion, they equilibrate quickly
to the local temperature and relative humidity conditions
by water evaporation. As evaporation is a molecular process, droplet
shrinking occurs very rapidly, see, for example Refs.~\cite{Dichotomy,ReidMicrophysics,Mastorakos2021,Boies2022},
and the droplet diameter after equilibration
is the droplet diameter most often experimentally accessible. We refer to the droplet
diameter at generation as the pre-evaporation diameter,  $\diPre$, and
that after equilibration as the post-evaporation diameter, $\diPost$. 
Their ratio defines the evaporation factor~\cite{Editorial2} $\zevap$
\beq
\diPost = \zevap \diPre .
\label{eq:zevap}
\eeq
The pre-evaporation diameter, via  $\rho_p$ the pathogen concentration at the 
location of droplet generation, e.g., oral region, determines the number of pathogens
$\Npath$, within a $\diPre$ droplet,
\beq
\Npath = \rho_p (\diPre) \times\frac{\pi}{6} \big ( \diPre \big )^3 =
\rho_p (\diPre) \times\frac{\pi}{6} \big ( \diPost /\zevap \big )^3 .
\label{eq:Npath}
\eeq
The post-evaporation diameter determines the physical properties
of the droplet like the removal rate $\lambdaDep$, via gravitational settling or other
surface-deposition processes, and droplet transport processes.
We also consider that it determines their lung-deposition probability $q_{d_i}$.
These observation confirm the importance of the evaporation factor $\zevap$,
a factor that depends strongly on the ambient relative humidity. Not only does it
determine $\Npath$ and the droplet deposition and transport properties,
it also influences viral infectivity, and eventually the viral inactivation
rate $\mu_{d,c}$ in that changes in the droplet diameter lead to changes of
the concentration of the within-droplet species~\cite{Editorial2}.
These concentration changes may have important consequences since, for example,
increased concentration of salts, proteins, organics, and acid may
damage the pathogen and modify its infectivity~\cite{WalterNenes2023,ReidBasic2022}.

\subsubsection{Transmission rates}

The infection transmission rates depend  on numerous parameters that may be categorised as
biological, behavioral, or physical.  In Ref.~\cite{NikosYD2010} we showed that 
the transmission rate associated with a $\diPost$ droplet,  be it airborne $\beta_i^d$
or settled $\beta_i^c$, may be expressed as
\begin{subequations}
\begin{eqnarray}
\beta_i^d & =&  c \tau_{d_i}
\times \frac{B}{ V_{cl}} q_{d_i} \times p_d
\times \rho_p^{(i)} (d_i) \times\frac{\pi}{6} \big ( d_i^{\textrm{post}}/\zeta_{\textrm{evap}} \big )^3 
\times \epsilon_i^d ,  \quad i = 1, \ldots, \imax , 
\quad {\mbox{\textrm{airborne droplets,}}} \label{eq:betaAirborne} \\
\beta_i^c & =&  c \tau_{c_i}
\times \eta_c q_{c_i} \times p_c \times
\rho_p^{(i)} (d_i) \times\frac{\pi}{6} \big ( d_i^{\textrm{post}}/\zeta_{\textrm{evap}} \big )^3
\times \epsilon_i^c ,  \quad i = 1, \ldots, \imax, 
\quad {\mbox{\textrm{settled droplets,}}}
\end{eqnarray}
\label{eq:TransmissionRates}
\end{subequations}
where $\imax$ is the total number of droplet compartments as specified by their
post-evaporation diameter.
In the main text, we also argued that the breathing rate $B$ (m$^3$/day)
may be factored out in Eq.~(\ref{eq:betaAirborne})
to define 
\beq
\tilde{\beta}_i^d \equiv \frac{\beta_i^d}{B} = c \tau_{d_i}
\times \frac{1}{ V_{cl}} q_{d_i} \times p_d
\times \rho_p^{(i)} (d_i) \times\frac{\pi}{6} \big ( d_i^{\textrm{post}}/\zeta_{\textrm{evap}} \big )^3 \times \epsilon_i^d ,
\label{eq:betaTilde}
\eeq
a parameter that converts infectious respiratory droplets to infection quanta.

The parameters  in Eqs.~(\ref{eq:TransmissionRates}) that depend on biological
properties are: the pathogen concentration $\rho_p^{(i)}$ at the generation location
of droplet $\diPre$
(number per volume) which
we take to be droplet-size dependent; 
the probability of infection $p_d$ due to
a lung-deposited airborne droplet per pathogen (dimensionless); and
the probability of infection $p_c$ due to
a settled droplet that has been transferred from a surface to a susceptible individual
facial membranes per pathogen. The breathing rate $B$ may also be consider
a biological parameter, but we prefer to consider it a physical parameter (see Table~\ref{table:SimParameters}).
Lastly, the infection recovery rate $\mu_I$ (number per day), not present in
Eqs.~(\ref{eq:TransmissionRates}), is also a biologically-determined parameter.

We consider the lung-deposition probability $q_{d_i}$ of a $\diPost$ droplet to be
a physically determined parameter. The characteristic personal-cloud volume, the
volume surrounding an individual, is denoted
by $V_{cl}$. Recently, Xenakis (2023)~\cite{xenakis2023} referred to the personal-cloud volume
as the ``breathing zone volume, i.e.,  the air volume surrounding a susceptible occupant
and determining their epidemiological status".

The transmission-rate parameters that depend on an individual's behavior include the
individual-infectious person average contact rate $c$ (number per day), and the transfer rate of
settled droplets to facial mucus membranes $\eta_c$ (number per day).  During each
infectious-susceptible encounter,
the susceptible individual is exposed to airborne  infectious droplets for a droplet-depending
breathing time $\tau_{d_i}$ (days), and to settled infectious droplets for the  duration of
a hands-face exposure time $\tau_{c_i}$ (days). The combination of these
average exposure times per contact leads to an average total exposure time to infectious
droplets per day of
$\langle \tau_{\textrm{exp}} \rangle = c \times \sum_i^{i_{\textrm{max}}}
 (\tau_{d_i}+ \tau_{c_i})$.
 
 The parameters $\epsilon_i^{d,c}$ include other effects that could modify the
 transmission rates and not initially considered in Ref.~\cite{NikosYD2010}. For example,
 the filtration efficiency of personal protective equipment or face masks 
 is an important factor that should be included in $\epsilon_i^d$. 

\subsubsection{Removal rates}

The droplet removal rates are \emph{effective} removal rates of infectious
droplets in that they include virus inactivation in addition to more traditional removal rates
like surface deposition or removal induced by indoor air ventilation. The removal
rates of airborne $\alpha_i^d$  and settled $\alpha_i^c$ droplets of post-evaporation
diameter $\diPost$ are
\begin{eqnarray}
\alpha_i^d & = &  \big ( 1 + c \tau_{d_i} \big ) \frac{B}{V_{cl}} q_{d_i} +
\mu_d + \lambdaDep^i (\diPost) + \lambda_{\textrm{air}} + \phi_i^d ,  \quad i = 1, \ldots, \imax,
\quad \mbox{\textrm{airborne droplets,}} \label{eq:RemovalRatesD} \\
\alpha_i^c & = &  \big( 1 + c \tau_{c_i} \big ) \eta_c q_{c_i} + \mu_c + \phi_i^c , 
\quad i = 1, \ldots, \imax, \quad {\mbox{\textrm{settled droplets}}}.
\label{eq:RemovalRatesC}
\end{eqnarray}
Similarly to the infection transmission rates, droplet removal mechanisms
may be associated with behavioural, biological, or physical processes.
The first term in both Eq.~(\ref{eq:RemovalRatesD}) and (\ref{eq:RemovalRatesC})
is a self removal term: in the case of airborne droplets it models removal by  inhalation by
the susceptible (shown to be negligible for influenza-related
parameters~\cite{MargueriteJTB}), in the case of settled droplets is self-transfer of a deposited droplet to
facial membranes.The viral inactivation rate in airborne droplets is denoted
by $\mu_d$ (number per day), and that of settled droplets by $\mu_c$
(number per day). They are determined by the properties
of the virus under ambient conditions, and hence a strong function of the relative
humidity~\cite{Editorial2}. The ventilation rate is denoted by $\lambdaVent$ (number of air
exchanges per day), whereas the surface deposition rate is denoted by $\lambdaDep$ (number of
droplets per day.)  In our simulations we considered that the only
physical processes that leads to droplet deposition is gravitational
settling, $\lambdaDep^i = \theta(\diPost)$. 

The parameters $\phi_i^{d,c}$ denote
any other process that might induce particle removal: for example, UV radiation
would be an additional viral inactivation mechanism that would modify $\mu_{d,c}$. Another
possible inactivation mechanism would be indoor spraying 
of nonhazardous levels of an acid, e.g., nitric, to decrease
droplet pH~\cite{WalterNenes2023} or spraying a basic solution to increase indoor micro-environmental
conditions to basic~\cite{ReidBasic2022}.

\subsubsection{Droplet generation rates}

Normal oro-nasal activities, like breathing, talking, laughing, singing, and more
violent expiratory events, like sneezing and coughing, produce a distribution
of respiratory droplet sizes.  As we try to retain features of the spread of SARS-CoV-2 we
opted to limit the estimate of the droplet generation rates to normal oro-nasal activities.
In addition, we neglect super-spreaders, or super-emitters,~\cite{SuperSpreaders2019}.
The generation rates
we analyzed are based on measurements reported by Johnson et al. (2011)~\cite{Johnson2011},
see, also, de Oliveira et al. (2021)~\cite{Mastorakos2021} and
Stettler et al.~\cite{Boies2022}.
We used the first two distributions~\cite{Johnson2011}, B (bronchiolar droplet generation mode)
and L (laryngeal droplet generation mode), to determine
the concentration-weighed droplet diameter $\SmallDiam$. Their emission rate was determined
from the reported data for $\textrm{Cn}_i$, the droplet number concentration (number of
droplet per cm$^3$). The droplet
concentration was converted to droplet number per second via the flow rate
of the Aerodynamic Particle Sizer (APS) of $5$ lt/min. The emitted respiratory droplet
per second was converted to number of expelled droplets per day by assuming 1.5 hours
of speaking per day (hence the explicit 1.5 in Table~\ref{table:SimParameters}).

Since the APS measures aerosol particles in the size range $0.50 \leq d_p \leq 20$,
we decided to use the data of Ref.~\cite{Johnson2011} only for the the smaller diameter $\SmallDiam$.
The emission rate of the $\LargeDiam$ droplets was based on the data of
Loudon and Roberts (1967)~\cite{LoudonRoberts1967}, as described in Ref.~\cite{NikosYD2010},
and preserving the total volume of the expelled oral fluid.



\begin{table}[!htb]
\centering
\caption{Simulation parameters: two airborne droplet classes.}
\label{table:Compare}
\begin{ruledtabular}
\begin{tabular}{cccc}
Parameter & Description & Estimate & Reference \\ \hline
\multicolumn{4}{l}{Biological Parameters} \\
$\rho_p^{(1)}$ & pathogen & $7.0 \times 10^7$ \#/cm$^3$ & Stadnytskyi et al. (2020)~\cite{Bax2020} \\
& concentration ($\SmallDiam$) & (viral copies /cm$^3$) &  \\

$\rho_p^{(2)}$ & pathogen & $3.50 \times 10^6$ \#/cm$^3$ & \textit{ibid.}  \\
& concentration ($\LargeDiam$)& (viral copies /cm$^3$)  & \\

$\mu_I$ & infection & $0.1677 $  &  Kevrekidis et al. (2021)~\cite{PGK-Greece} \\
& recovery rate & (per day) & \\

$\mu_d$ & inactivation & $15.13 $  & van Doremalen et al. (2020)~\cite{vanDoremalen},  \\
& rate (airborne) & (per day) & Buonanno et al. (2020)~\cite{Buonanno2020a}  \\

$p_d$  & probability of & 0.052  & Drossinos and Stilianakis (2010)~\cite{NikosYD2010} \\
&  infection (airborne) & (-) &  \\

$1/ \sigma$ & latent period & 0.1 or 6 days & Scenario parameter \\

\multicolumn{4}{l}{Behavioural Parameters} \\
$c$ & contact rate & 18 \#/day  & Sypsa et al.  (2021)~\cite{Sypsa2021} \\
& per day & & \\
$\tau_{d_1}$ & characteristic breathing  & 25 min & Based on \\
& time ($\SmallDiam$) & &  Drossinos and Stilianakis (2010)~\cite{NikosYD2010}   \\
$\tau_{d_2}$ & characteristic breathing & 1 min & \textit{ibid.}  \\
& time $(\LargeDiam$) & & \\

\multicolumn{4}{l}{Physical and physiological parameters} \\
$\SmallDiam$ & small-droplet diameter & 2.05 $\mu$m  & Johnson et al. (2011)~\cite{Johnson2011} \\
& speaking & & \\
$\LargeDiam$  & large-droplet diameter & 82.13 $\mu$m & Loudon and Roberts (1967)~\cite{LoudonRoberts1967} \\
& speaking & & \\
$\zeta_{\textrm{evap}}$  & evaporation factor & 0.40 (-)  &  Lieber et al. (2021)~\cite{EvapLieber} \\
$B$ & breathing rate & 12 m$^3$/day & Drossinos and Housiadas (2006~\cite{MultiphaseFlow}  \\
$V_{cl}$  & volume personal cloud & 8 m$^3$ &  Drossinos and Stilianakis (2010)~\cite{NikosYD2010} \\
$q_{d_1}$ & inhaled-droplet & 0.88 (-) &  Drossinos and Housiadas (2010)~\cite{MultiphaseFlow}\\
&  deposition probability ($\SmallDiam$) & \\
$q_{d_2}$ & inhaled-droplet & 1.00 (-) & \textit{ibid.} \\
&  deposition probability ($\LargeDiam$) & & \\

$\kappa_1^d$  & airborne droplet generation rate &  $1.5 \times 51,182 = $  & Johnson et al. (2011)~\cite{Johnson2011} \\
& speaking (droplets/day)($\SmallDiam$) & $76,773$ \#/day & \\

$\kappa_2^d$ & airborne droplet generation rate & $0.8 \times 47,160= $ & Loudon and Roberts (1976)~\cite{LoudonRoberts1967} \\
& speaking (droplets/day) ($\LargeDiam$) &  $37,728$ \#/day  & \\
$\lambdaDep^1 = \theta_1 $  & airbone droplet deposition rate &  $7.8$  \#/day & Drossinos and Housiadas (2006)~\cite{MultiphaseFlow} \\
& still-air gravitational settling ($\SmallDiam$) & & \\
$\lambdaDep^2 = \theta_2$ & settled droplet generation rate & $10,558$ \#/day &  \textit{ibid.}  \\
& still-air gravitational settling ($\LargeDiam$) &  & \\

$\lambda_{\textrm{air}}$ & air exchange rate (AER) & 4.8 exchanges/day &   Buonanno et al. (2020)~\cite{Buonanno2020a} \\
& & &  \\

\multicolumn{4}{l}{Infection-risk parameter} \\
$\deltat$ & prediction time & 1 or 7 days & estimate \\
\end{tabular}
\label{table:SimParameters}
\end{ruledtabular}
\end{table}

\subsubsection{Other parameters}

All simulation parameters, along with the associated references, are reported
in Table~\ref{table:SimParameters}. 
We note that observations~\cite{Ventilation2020} and
simulations~\cite{Editorial2} suggest the importance of the
ventilation rate. We chose to use a characteristic value for
typical Italian buildings as reported in Ref.~\cite{Buonanno2020a},
namely $\lambdaVent = 0.2$ air exchanges per hour.
%
The evaporation factor $\zevap$ was chosen to be $0.40$,
an intermediate value between the recent estimate~\cite{EvapLieber}
of $0.20$ and our initial estimate~\cite{NikosYD2010} of $0.50$.
The viral inactivation rate in airborne droplets was based on the
early measurements of van Doremalen et al. (2020)~\cite{vanDoremalen}.
It is frequently quoted~\cite{Buonanno2020a, Buonanno2020b} as the removal rate  in
terms of the viral half-life $t_{1/2}$ as  $\lambda_{\textrm{inact}} = \ln(2)/t_{1/2}$:
both references used $\lambda_{\textrm{inact}} = 0.63$ per hour gives $\mu_d = 38$ per day.  Our estimate is slightly smaller.


%

\end{document}